\documentclass[%
 reprint,
 amsmath,amssymb,
prb,
]{revtex4-1}

\usepackage{graphicx}
\usepackage{dcolumn}
\usepackage{bm}
\usepackage{hyperref}
\usepackage{bbold}
\usepackage[caption=false]{subfig}
\usepackage{bm}
\newcommand{\vect}[1]{\boldsymbol{\mathbf{#1}}}
\usepackage{setspace}
\usepackage{cleveref}

\usepackage{float}
\usepackage{paralist}

\usepackage{tikz}
\usepackage{tikz-3dplot}
\usetikzlibrary{3d,shadings, fadings}
\usetikzlibrary{shapes.geometric, arrows}
\tikzstyle{arrow} = [thick,->,>=stealth]
\tikzfading[name=fade out, inner color=transparent!0, outer color=transparent!100]
\usetikzlibrary{arrows,positioning} 
\tikzset{
    >=stealth',
    punkt/.style={
           rectangle,
           rounded corners,
           draw=black, very thick,
           text width=6.5em,
           minimum height=2em,
           text centered},
    pil/.style={
           ->,
           thick,
           shorten <=2pt,
           shorten >=2pt,}
}

\begin{document}


\title{Effective model for Majorana modes in graphene}

\author{A. L. R. Manesco \thanks{A.M. is the corresponding author.}}
 \email[Corresponding author:\ ]{antoniolrm@usp.br}
\affiliation{Lorena Engineering School, University of S\~{a}o Paulo.}
\author{G. Weber}%
\affiliation{Lorena Engineering School, University of S\~{a}o Paulo.}%
\author{D. Rodrigues Jr}
\affiliation{Lorena Engineering School, University of S\~{a}o Paulo.}





\begin{abstract}
It was recently proposed that the interface between a graphene nanoribbon in the canted antiferromagnetic quantum Hall state and a s-wave superconductor may present topological superconductivity, resulting in the appearance of Majorana zero modes.\cite{Lado2015Majorana} However, a description of the low-energy physics in terms of experimentally controllable parameters was still missing. Starting from a mean-field continuum model for graphene in proximity to a superconductor, we derive the low-energy effective Hamiltonian describing the interface of this heterojunction from first principles. A comparison between tight-binding simulations and analytical calculations with effective masses suggests that normal reflections at the interface must be considered in order to fully describe the low-energy physics.

\end{abstract}

\pacs{Valid PACS appear here}
\maketitle


\section{Introduction}

Zero-energy excitations in symmetry-protected topological superconducting systems are predicted to behave as zero energy Majorana quasi-particles.\cite{Leijnse2012Introduction, Beenakker2013Search, Elliot2015Colloquium, Sato2017Topological,aguado2017majorana} Despite the recent experimental efforts to capture signatures of such excitations, there is still no general consensus regarding the existence of these ellusive zero-energy modes.\cite{He2017Chiral, Zhang2018Quantized, Banerjee2018Observation, Kasahara2018Majorana} This problem stimulated a plethora of theoretical proposals for systems supporting Majorana modes in a variety of nanosystems, from nanowires to two-dimensional heterostructures.\cite{Kitaev2001Unpaired, Fu2009Josephson, Akhmerov2009Electrically, Stanescu2013Majorana, Dumitrescu2015Majorana, Sato2017Topological} In this paper, we consider the proposal of one-dimensional topological superconductivity at the interface of graphene/superconductor junctions,\cite{Lado2015Majorana} experimentally motivated by ballistic junctions in quantum Hall regime and tunability of magnetic ordering.\cite{Young2012Spin, Young2014Tunable, Calado2015Ballistic, Shalom2016Quantum}

\begin{figure}\label{scheme}
\subfloat[\label{CAFM}]{\includegraphics[width=0.25\textwidth]{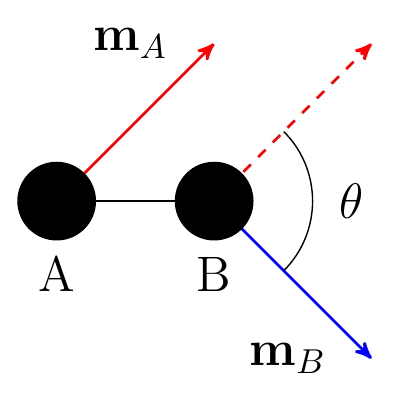}}\\
\subfloat[\label{setup}]{\includegraphics[width=0.5\textwidth]{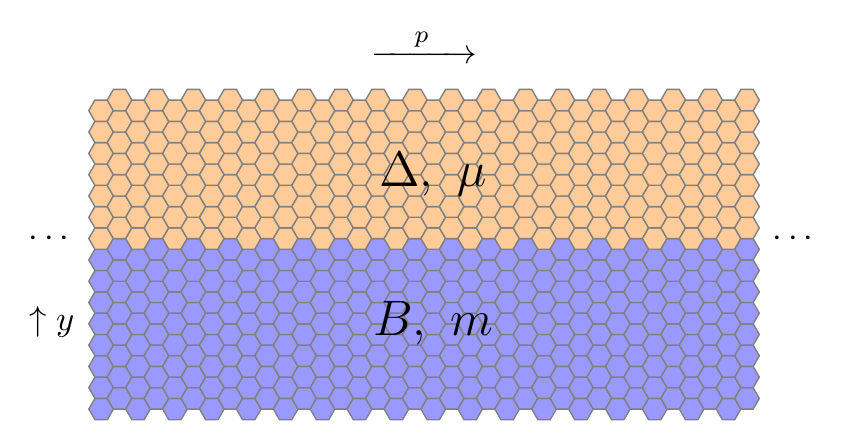}}
\caption{(a) Scheme of the magnetization in the $A$ and $B$ sublattices; $\theta$ denotes the canting angle between both magnetizations. (b) Schematic representation of the graphene-superconductor setup considered in this work. For the derivation of the effective model, we considered an infinite ribbon along the $x$-axis (so that $k_x$ is a good quantum number). For negative values of $y$ (blue region), we added an orbital magnetic field, $B$, to induce Landau levels. The emergent magnetic ordering, with magnitude $m$, was treated as a mean field term. For $y>0$ (orange region), we included an induced $s$-wave superconducting order parameter, $\Delta$, and chemical potential, $\mu$, as a result of a $s$-wave superconductor deposited over this region. The continuum Hamiltonian for this system is written in \eqref{hardwall}.}
\end{figure}

The possible appearance of Majorana modes in graphene relies on the interplay between three different phenomena (scheme in Fig. \ref{scheme}).\cite{Lado2015Majorana} First, each of the two degenerate zero energy eigenstates in the zeroth Landau level of quantum Hall graphene is restricted to a distinct combination of valley and sublattice indices. Hence, there is an intrinsic identification of sublattice and valley degrees of freedom.\cite{Goerbig2011quantum, Bernevig2007quantum, Lado2015Edge} Second, electronic interactions lead to the emergence of a canted antiferromagnetic ordering that can be tuned by an applied Zeeman field, as shown in Fig. \ref{CAFM}.\cite{Young2012Spin, Kharitonov2012Edge, Young2014Tunable, Lado2014Noncollinear} Thus, the aforementioned identification is enlarged to include spin degrees of freedom as well. As consequence, counter-propagating edge states with different helicities emerge, just as in a quantum spin Hall insulator (QSHI).\cite{Kane2005Quantum, Lado2015Majorana} Finally, by inducing a superconducting order parameter the system becomes gapped and Majorana zero modes may emerge.\cite{Fu2009Josephson} We also note that a similar phenomenon was predicted to occur at the interface between  superconductors and antiferromagnetic insulators,\cite{PhysRevLett.121.037002} suggesting that the role of the quantum Hall state in graphene is just to induce high correlations in the flat-band zeroth Landau level.\cite{Lado2014Noncollinear}

Mean-field simulations of such graphene/superconductor junctions  corroborated this proposal for hosting Majorana zero modes, and an \emph{ad hoc} phenomenological model for the edge states was proposed from numerical band diagram calculations. The model is described by the following effective low-energy Hamitonian:\cite{Lado2015Majorana}
\begin{align}\label{hlado}
H_{\text{eff}}=\left(
\begin{array}{cccc}
\mu_1 + v_1 p & b_{\theta} & w & 0 \\
b_{\theta} & \mu_2 - v_2 p & 0 & w \\
w & 0 & - \mu_2 - v_2 p & b_{\theta}\\
0 & w & b_{\theta} & -\mu_1 + v_1 p
\end{array}\right),
\end{align}
where $\mu_i$ and $v_i$, $i=1,2$, are on-site energies and propagation velocities of the chiral modes, respectively. The topological gap is denoted by $b_{\theta}\propto \cos \theta$, where $\theta$ is the canting angle between the magnetic moment of each sublattice, see Fig \ref{CAFM}. The ferromagnetic state corresponds to $\theta =0$, while the antiferromagnetic state to $\theta=\pi$. The parameter $w$ represents the intervalley coupling and $p$ is the quasi-momentum along the interface direction. The Hamiltonian \eqref{hlado} is in the same representation of \eqref{heff1}.

Starting from a mean-field continuum model for graphene, we derive the effective Hamiltonian \eqref{hlado} from a more constructive approach. Besides providing a better understanding of the physics described by \eqref{hlado}, our approach allows us to express the phenomenological parameters in terms of real, experimentally controllable ones. In particular, we uncover the important role played by normal reflections at the graphene-superconductor interface, which have to be properly considered in order to fully describe the low-energy Hamiltonian. 

Our paper is organized as follows. In Sec. \ref{model}, we propose a continuum Hamiltonian to describe a strip of graphene in the quantum Hall canted-antiferromagnetic (QHCAF) phase in proximity to an $s$-wave superconductor (SC). We then obtain an effective Hamiltonian to describe the interface physics by projecting the continuum Hamiltonian onto the zero energy modes at the QHCAF/SC interface. In Sec \ref{term}, we argue that some terms present in the Hamiltonian \eqref{hlado} can only be derived by considering termination-sensitive normal reflections at the QHCAF/SC interface. To account for these phenomena we consider effective masses that describe the boundary conditions at the interface. In Sec. \ref{majorana_modes}, we briefly discuss the appearance of Majorana modes and the resulting topological classification. Finally, in Sec. \ref{conclusions}, we summarize our results and discuss some points that must be considered in future works. 

\section{The model}
\label{model}

We start with a low-energy continuum Hamiltonian for the QHCAF/SC junction (illustrated in Fig. \ref{scheme}). For $y<0$, we consider graphene in the presence of a perpendicular orbital magnetic field ($B$), leading to the quantum Hall state. We also include a mean-field staggered magnetization energy ($m$), proposed as the explanation for the gap opening at graphene's zeroth Landau level.\cite{Young2012Spin, Kharitonov2012Edge, Young2014Tunable, Lado2014Noncollinear} For $y>0$, we also consider graphene, but in proximity to an $s$-wave superconductor, leading to an induced order parameter ($\Delta$) and shifting the chemical potential to $\mu$, defined with respect to the Dirac cone. The presence of superconductivity repels the orbital magnetic field and, therefore, no magnetization is expected. In the valley-symmetric representation,\cite{Akhmerov2008Boundary} the Dirac-Bogoliubov-de Gennes (DBdG) Hamiltonian reads (using $\hbar=v=e=1$)
\begin{align}\label{hardwall}
\mathcal{H} =& \Pi_x \Gamma_1 + \Pi_y \Gamma_2 \nonumber \\
+& m \left(\Gamma_3  \sin \frac{\theta}{2} + \Gamma_4 \cos \frac{\theta}{2} \right) \Theta(-y) \\
+& \left(\Delta \Gamma_5 -\mu \Gamma_0\right)\Theta(y) \nonumber,
\end{align}
where $\Pi_i=p_i + A_i$ is the canonical momentum in the presence of a magnetic field $B$, $v$ is the Fermi velocity and $\theta$ is the magnetization canting angle between graphene sublattices. The Heaviside step function is denoted by $\Theta(y)$. In order to preserve translational symmetry along the $x$-axis, we consider the Landau gauge $\vect{A}=(By,0,0)$ for $y<0$. Moreover, we set $\vect{A} = (0,0,0)$ for $y>0$ to ensure the continuity of the gauge field and properly account for the Meissner effect. We relegate the explicit expressions for the spinors and $\Gamma$-matrices to Appendix \ref{gamma}.

\begin{figure}
\subfloat[\label{ferro}]{\includegraphics[width=0.225\textwidth]{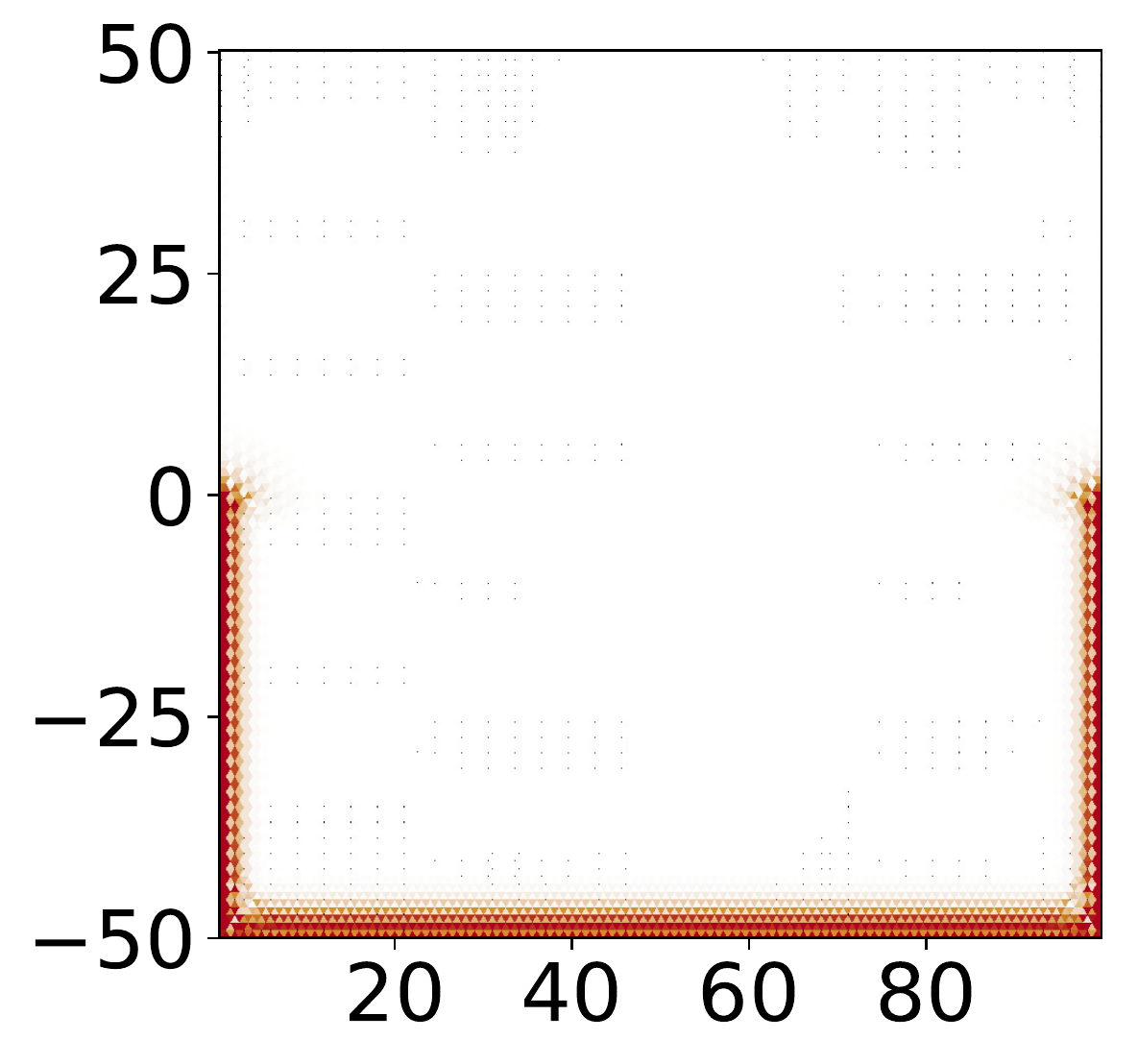}}
\subfloat[\label{antiferro}]{\includegraphics[width=0.225\textwidth]{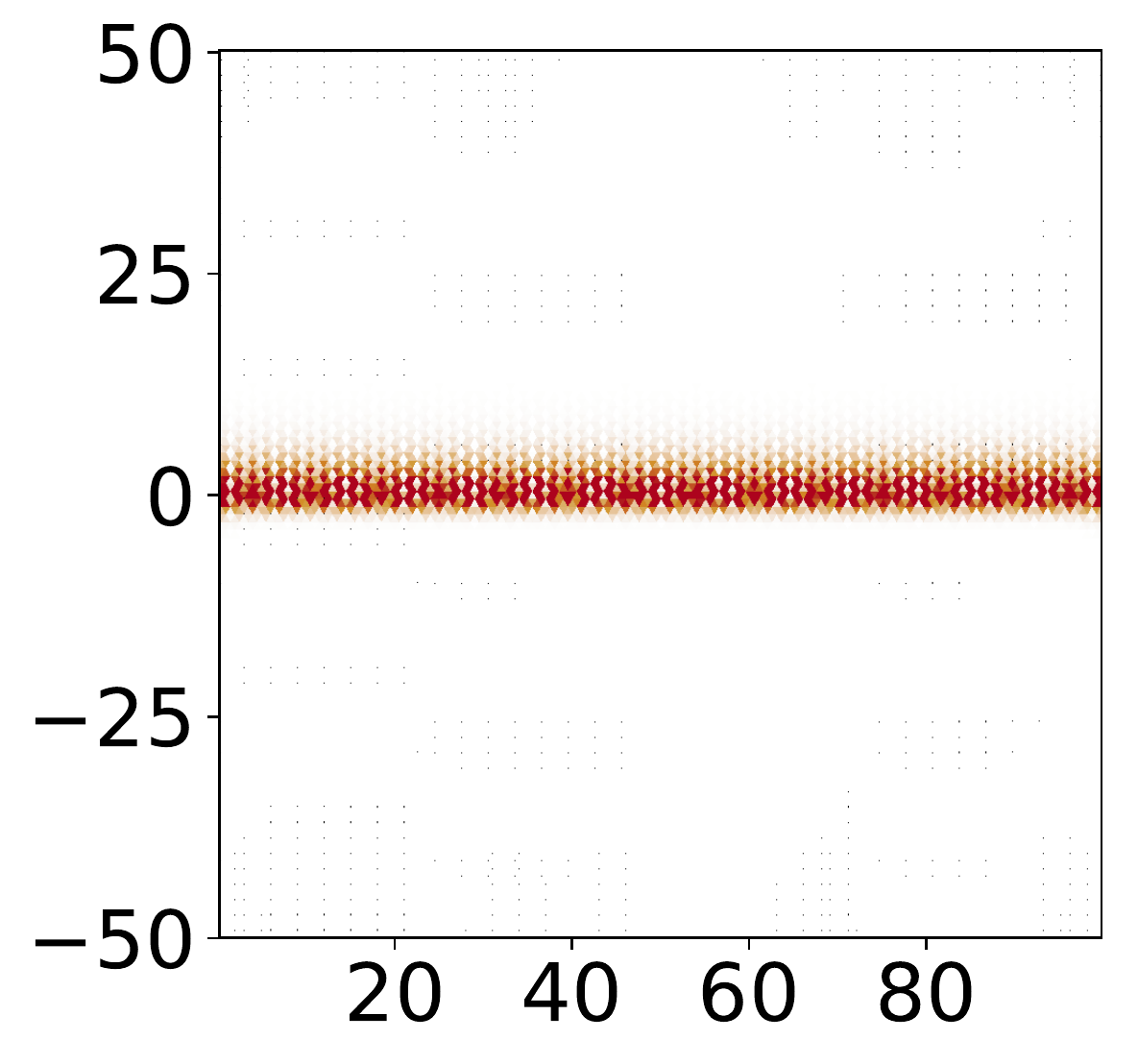}}\\
\subfloat[\label{canted}]{\includegraphics[width=0.225\textwidth]{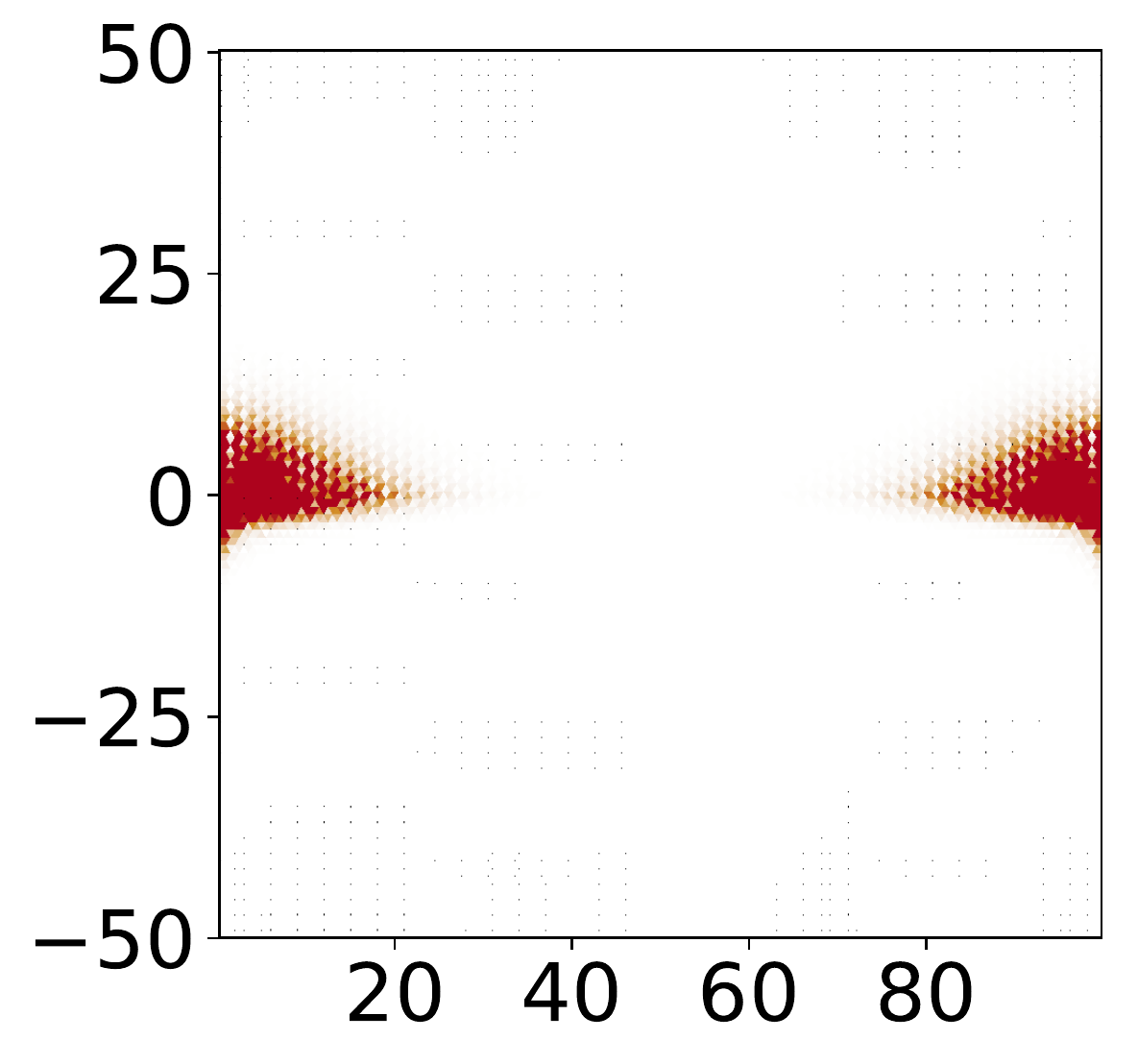}}
\subfloat[\label{majpol}]{\includegraphics[width=0.225\textwidth]{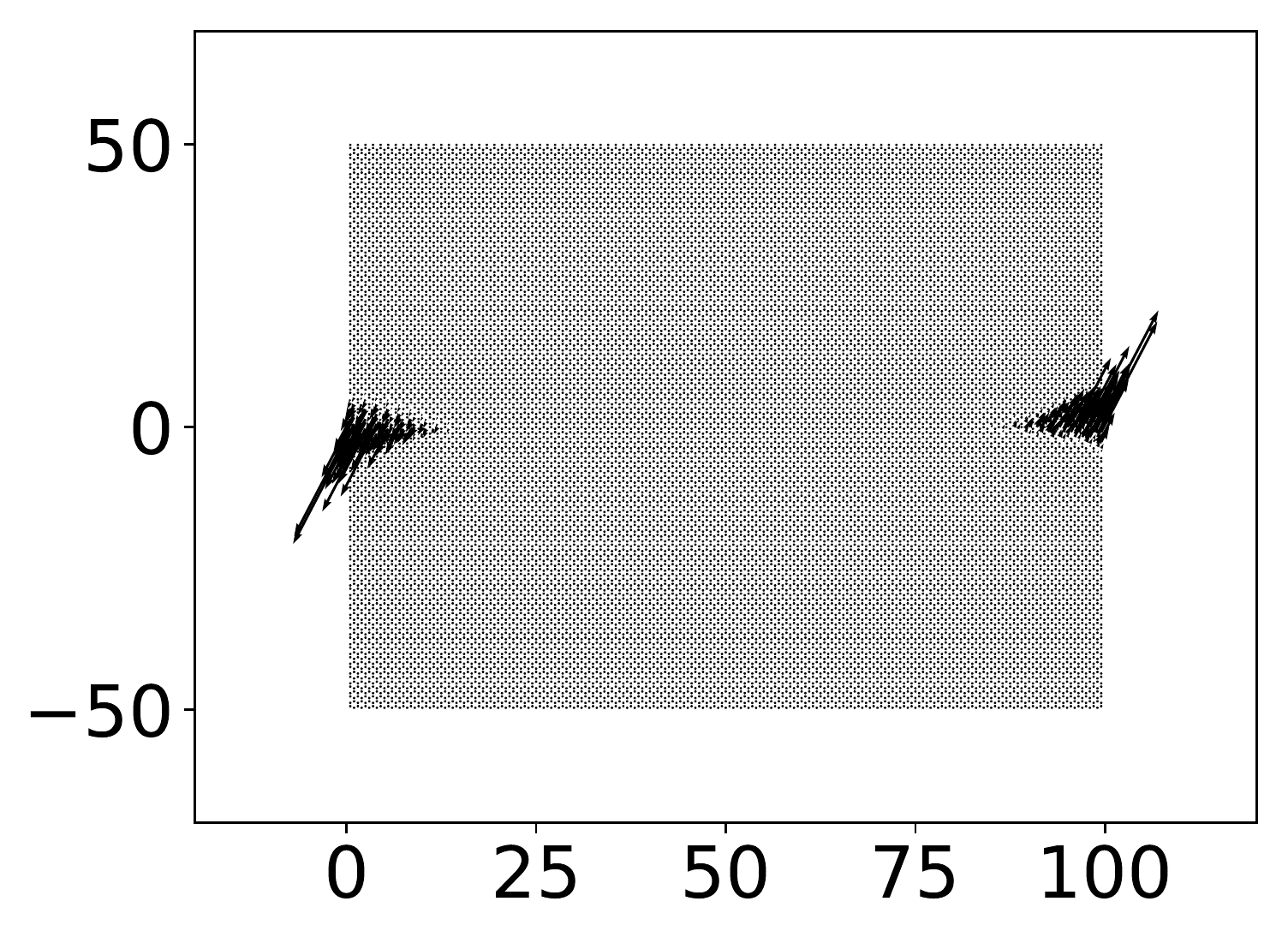}}
\caption{Local density of states and Majorana polarization for QHCAF/SC junctions. In the upper half of the panels, we have graphene with an induced $s$-wave superconducting coupling, whereas for the lower half we have graphene in: (a) ferromagnetic; (b) antiferromagnetic; (c) canted-antiferromagnetic ($\theta=\pi$) states. (d) Majorana polarization\cite{Bena2016Testing, Sticlet2012Spin} corresponding to (c). We used $B=0.05 \frac{\hbar}{ea^2}$, $m=0.5t$, $\mu=0.3t$ and $\Delta=0.25t$, where $t$ is the hopping energy in order to reproduce the phenomenology with lower computational cost. The length unit is the lattice constant.}
\label{LDOS}
\end{figure}

To properly integrate out the extra degrees of freedom and derive a model that describes only the interface states, we first examine numerical results from a tight-binding implementation of \eqref{hardwall} using the Kwant code.\cite{Groth2014Kwant} Figure \ref{LDOS} shows the local density of states and Majorana polarization\cite{Bena2016Testing, Sticlet2012Spin} at zero energy under three situations: ferromagnetic ($\theta=0$), antiferromagnetic ($\theta=\pi$) and canted-antiferromagnetic ($\theta=\pi/2$) orderings. Since only the antiferromagnetic state exhibits zero energy modes that extend all over the interface, we impose $\theta=\pi$ in \eqref{hardwall}. Setting, without any loss in generality, $p_x=0$, we solve $\mathcal{H}_0\psi=0$ to find the zero energy states of
\begin{align}\label{hardwalltheta=pi}
\mathcal{H}_0 =&  -i \Gamma_2 \partial_y \nonumber \\
+& \left(By \Gamma_1 + m \Gamma_3\right) \Theta(-y)\\
+&\left(\Delta \Gamma_5 -\mu \Gamma_0\right) \Theta(y). \nonumber 
\end{align}

Diagonalizing \eqref{hardwalltheta=pi}, we obtain the following eigenstates
\begin{align}\label{solutions}
\psi_{\alpha}(y) = \frac{1}{\mathcal{N}} e^{\lambda_{\alpha}(y)}\psi_{\alpha}^{(0)}(y),
\end{align}
where $\lambda_{\alpha}(y)$ and $\psi_{\alpha}^{(0)}(y)$ are, respectively, the eigenvalues and eigenspinors of
\begin{align}
\Lambda(y) =& -i \int_{0}^y d\xi\ \Gamma_2^{-1} \left[ \Theta(-\xi) (B \xi \Gamma_1 + m\Gamma_4) \right.\\
+& \left.\Theta(\xi)\left(\Delta \Gamma_5 -\mu \Gamma_0\right)\right],  \nonumber 
\end{align}
and $\mathcal{N}$ is a normalization constant. Next, we impose two physical constraints on the eigenfunctions: 
\begin{inparaenum}[(i)]
	\item regularity at spatial infinity, \emph{i.e.}, we discard all solutions $\psi_{\alpha}(y)$ that diverge as $y \to \pm \infty$;
	\item continuity at the interface.
\end{inparaenum}
We relegate the lengthy expressions for the resulting eigenbasis $\{\tilde{\psi}_{\alpha}\}$ to Appendix \ref{eigenfunctions}.

We can finally derive an effective Hamiltonian for general interface states by calculating
\begin{align}\label{expect}
\mathcal{H}_{\text{eff}}^{\alpha \beta} = \langle \tilde{\psi}_{\alpha}|\mathcal{H}-\mathcal{H}_0|\tilde{\psi}_{\beta}\rangle.
\end{align}
In terms of the spinor basis $\Psi=i(-\psi_{+-}, \psi_{++}, -\psi_{--}, \psi_{-+})^T$, it reads
\begin{align}\label{heff1}
\mathcal{H}_{\text{eff}} =\left(
\begin{array}{cccc}
\tilde{v} p & b_{\theta} & 0 & 0 \\
b_{\theta} & -\tilde{v} p & 0 & 0 \\
0 & 0 & -\tilde{v} p & b_{\theta}\\
0 & 0 & b_{\theta} & \tilde{v} p
\end{array}\right).
\end{align}
Thus, the effective degrees of freedom correspond to four chiral modes with the same propagation velocity $\tilde{v}$. There is also an intravalley coupling between the two different helicities, $b_{\theta}= \tilde{\Delta} \cos \theta$, that vanishes for antiferromagnetic ordering ($\theta=\pi$). The explicit expressions for $\tilde{v}$ and $\tilde{\Delta}$ can be found in \eqref{cumbersome}.

The resulting effective Hamiltonian \eqref{heff1} coincides with the phenomenological model \eqref{hlado} only in the case of vanishing on site energies, $\mu_1=\mu_2=0$, absence of intervalley coupling, $w=0$ and coinciding propagation velocities $v_1=v_2=\tilde{v}$. The first two deficiencies of our effective model are related to not taking into account the effect of terminations and normal reflections on the physics of the edge states. On the other hand, the indistinguishability of the propagation velocities is a limitation of the first order expansion performed near the Fermi level to obtain the continuum model for graphene. These issues will be dealt with in details in the next section. We note, nonetheless, that the topological gap, $b_{\theta}$, is the same in both models. Thus, one may expect the appearance of bound states whenever the gap changes sign, as explained in Sec. \ref{majorana_modes}.

\section{Effect of terminations and normal reflections}
\label{term}

It is well known that the effect of atomic structure in graphene-vacuum boundaries is crucial to fully describe the low-energy spectrum of finite systems.\cite{Akhmerov2008Boundary} However, to the best of our knowledge, so far no systematic study was performed to account for similar effects in graphene-superconductor junctions. In Sec. \ref{model}, the only boundary condition imposed at the interface was the continuity of the eigenspinors to enforce the effects of Andreev reflections.\cite{Beenakker2006Specular, Titov2006Josephson} Thus, termination physics associated with normal reflections were not considered. In the following, we provide numerical evidence that termination dependent boundary conditions corresponding to normal reflections are needed to fully describe the low-energy dynamics.

\subsection{Numerical analysis}

\begin{figure}[h!]
\center
\subfloat[\label{zigzag}]{\includegraphics[scale=0.25]{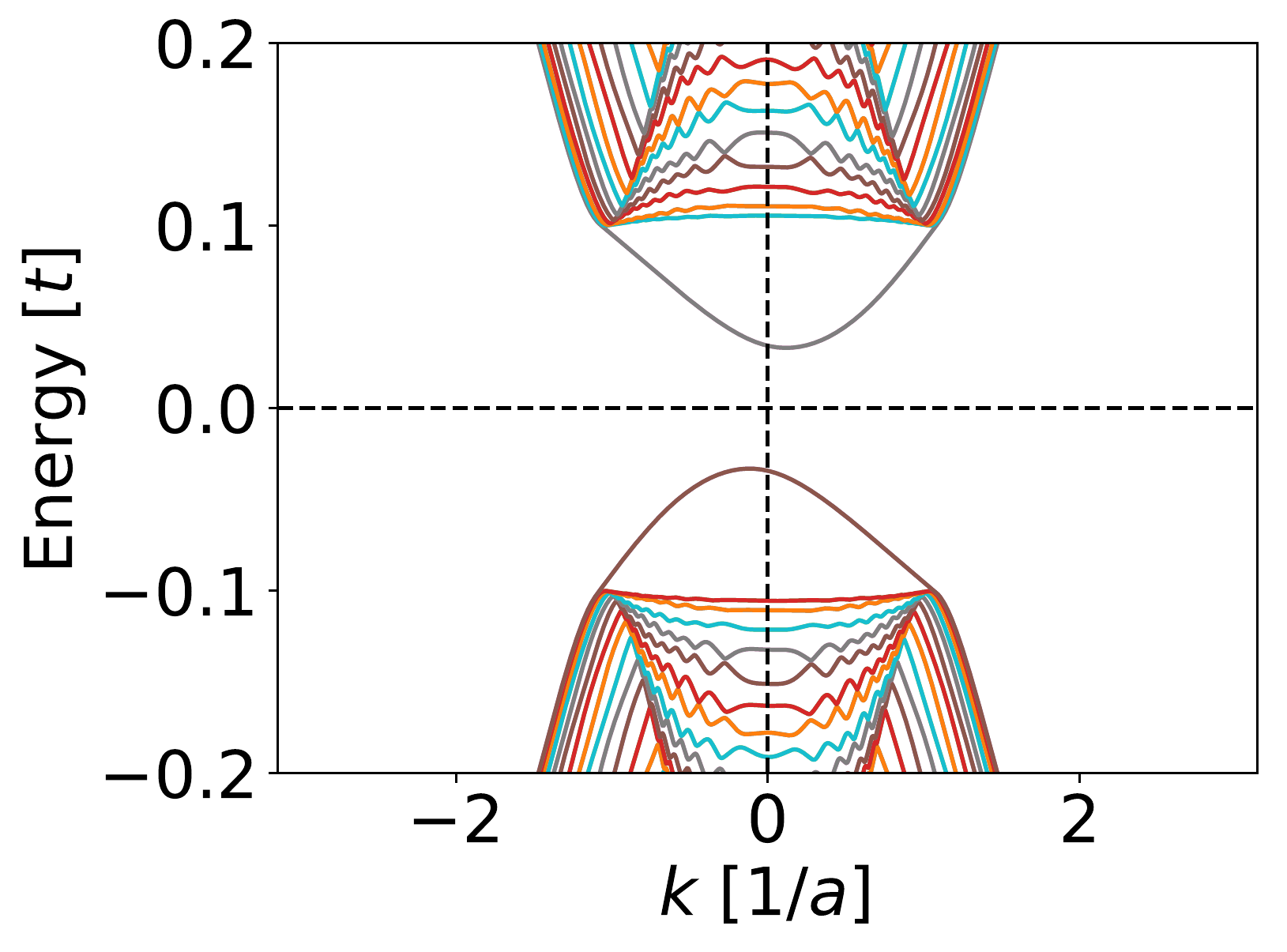}}
\subfloat[\label{armchair}]{\includegraphics[scale=0.25]{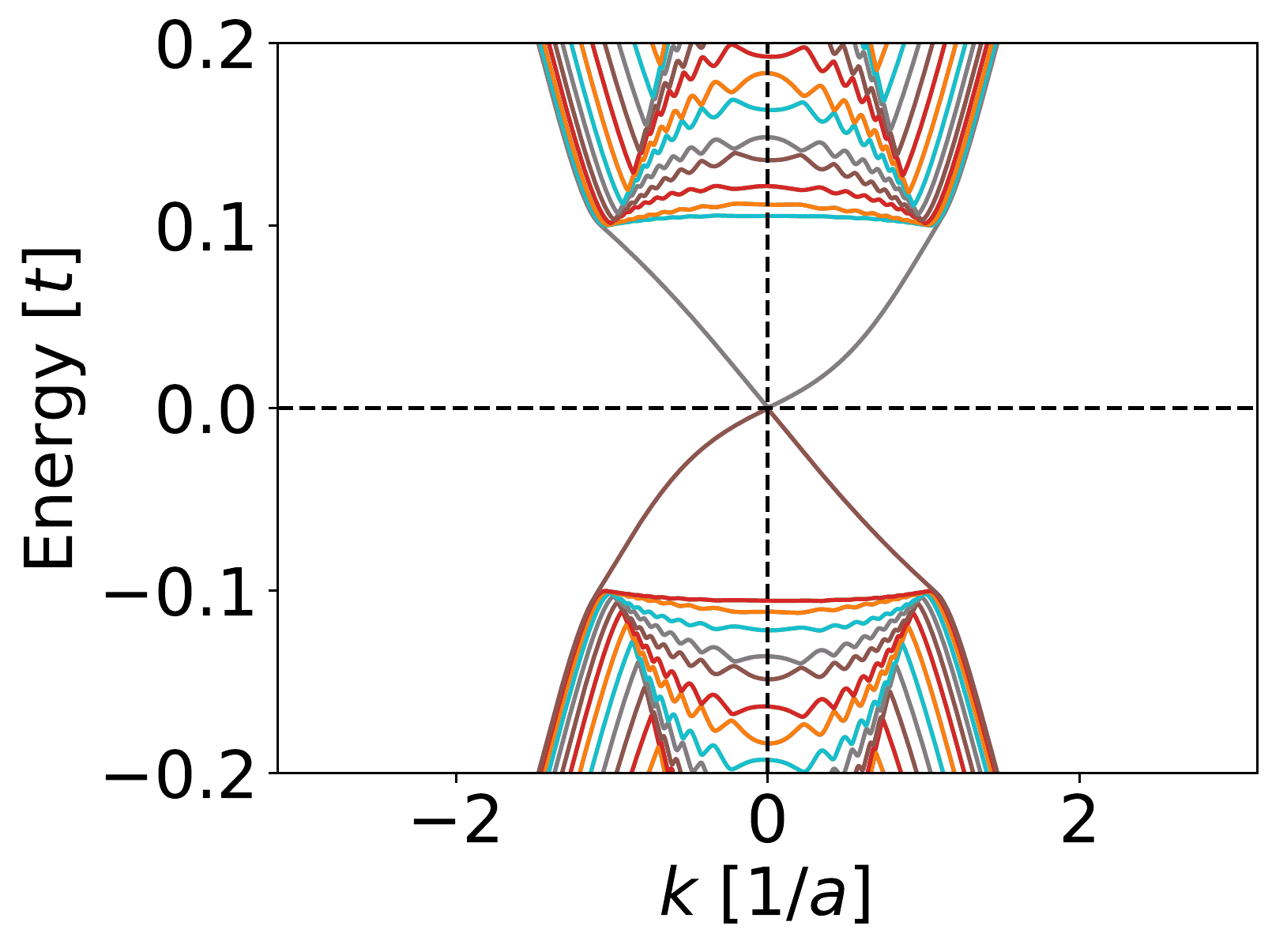}}\\
\subfloat[\label{armchair}]{\includegraphics[scale=0.25]{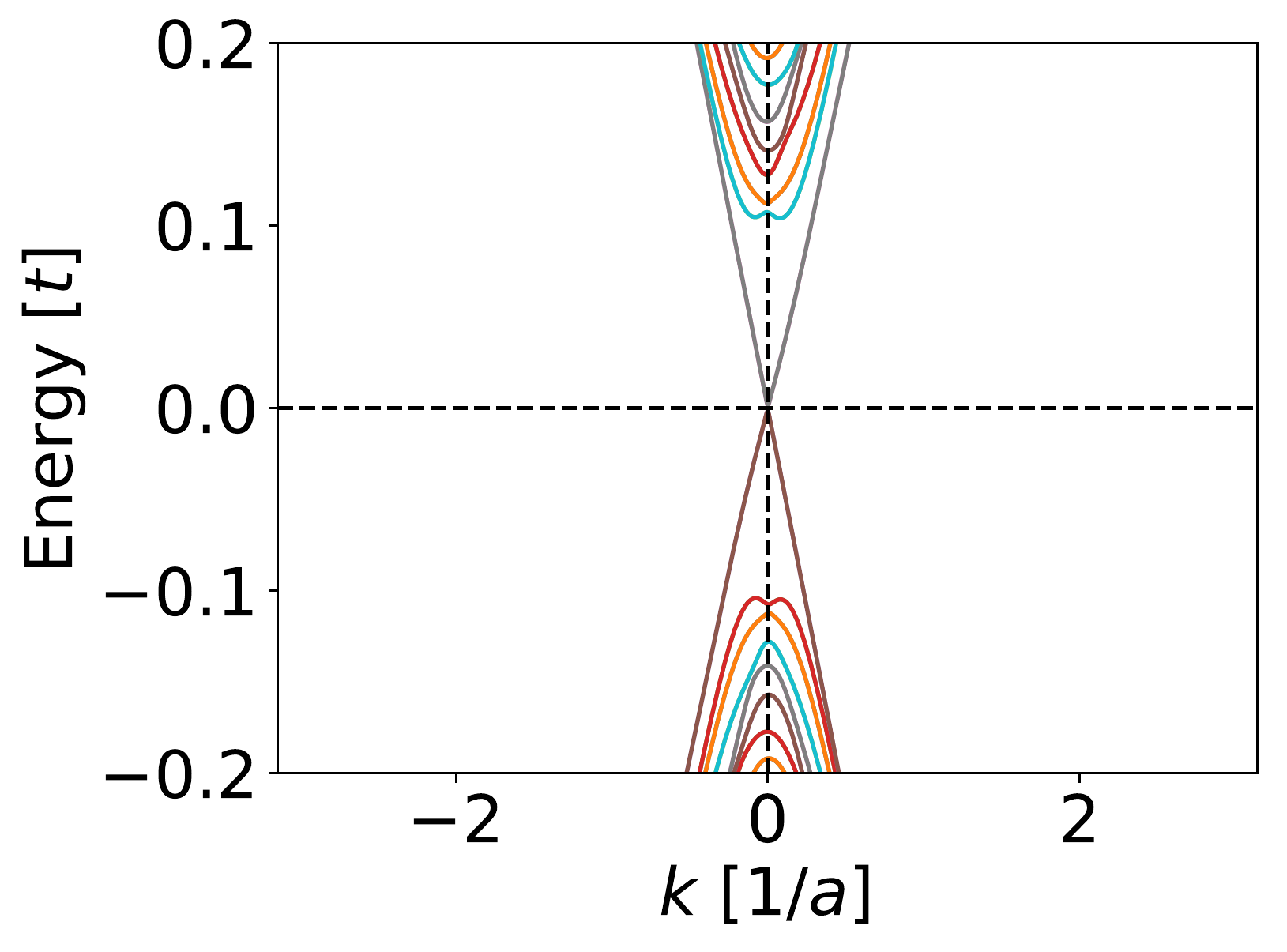}}
\subfloat[\label{armchair}]{\includegraphics[scale=0.25]{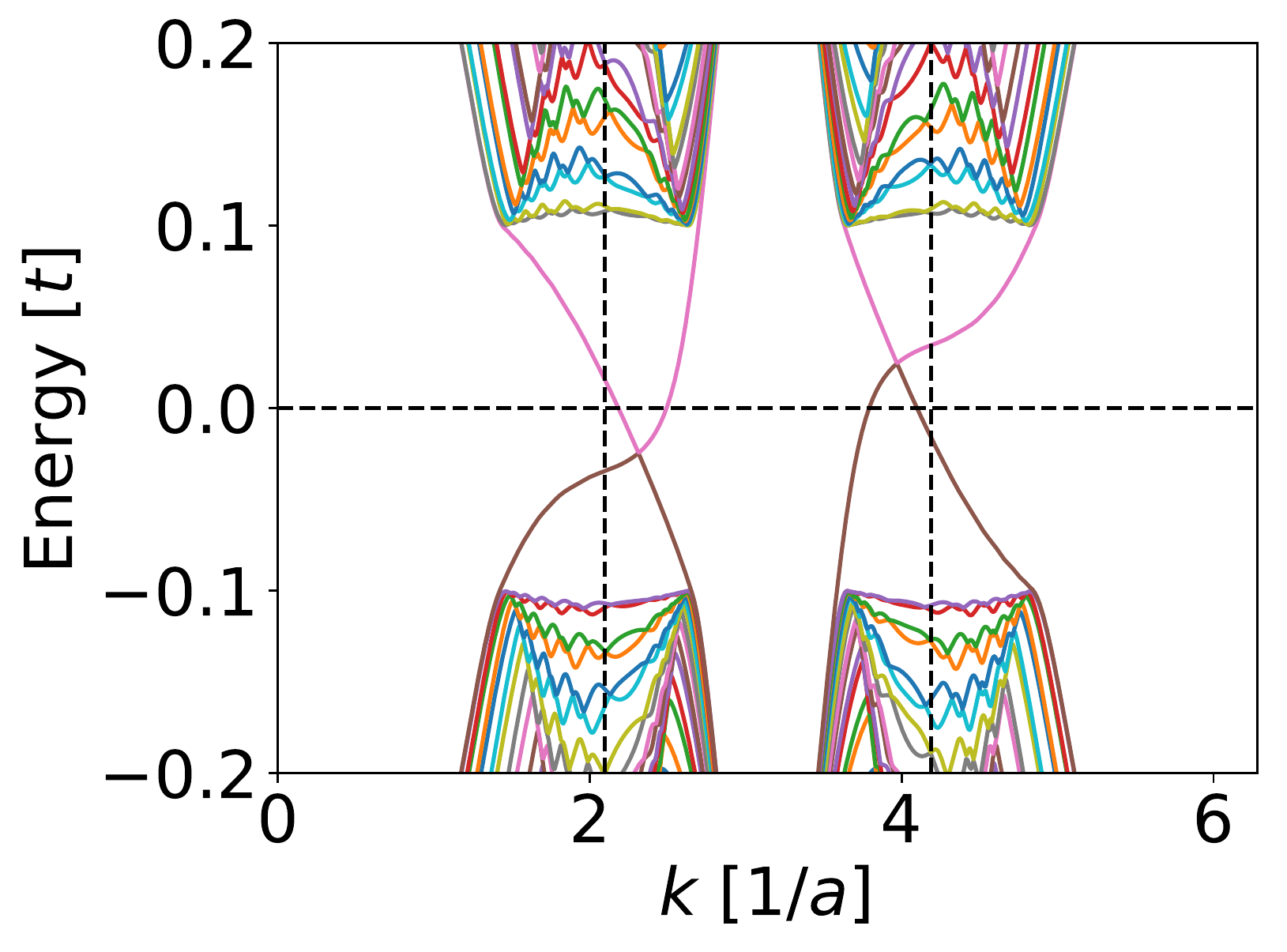}}\\
\subfloat[\label{armchair}]{\includegraphics[scale=0.25]{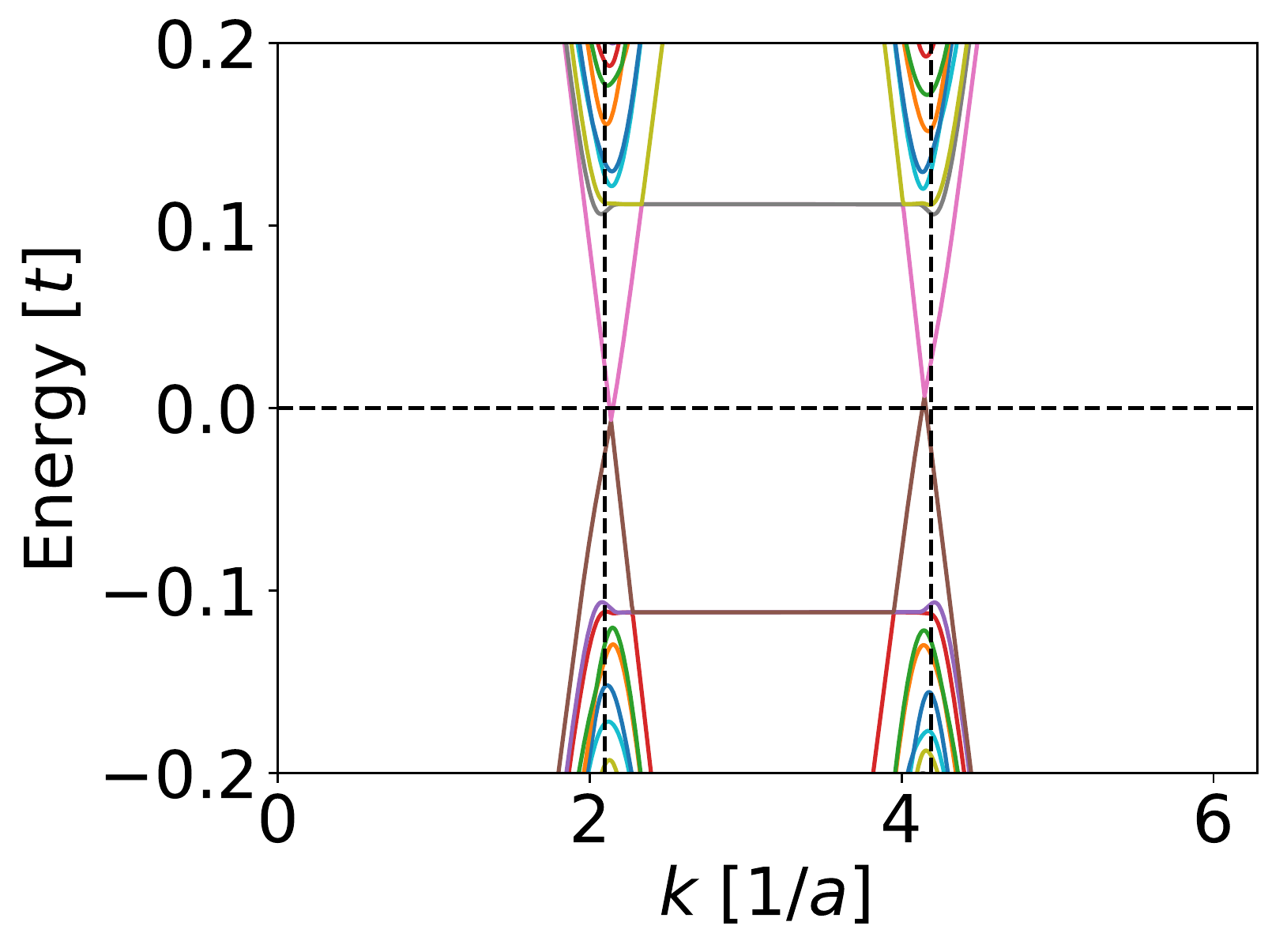}}
\caption{Band diagrams for pristine armchair (a), (b) and (c) and zigzag (d) and (e) graphene nanoribbons. All calculations were performed for $60a$ width ribbons divided in half: the superconducting-induced region for $y>0$ and quantum Hall regime for $y<0$. Also, $B=0.05 \frac{\hbar}{ea^2}$, $m=0.5t$ and $\Delta=0.1t$, where $t$ is the hopping constant. These parameters do not correspond to realistic conditions, but they preserve the phenomenology and have lower computational costs. For the armchair ribbons, (b) and (c), a smooth variation of the parameters were used, proportional to $\tanh y$, which results in a significant reduction of the gap, when compared with the variation proportional to the Heaviside step function (a). In (a), (b) and (d), $\mu=0.5t$, while, for (c) and (e), $\mu=0.05t$.
We stress that for higher values of $\mu$, there is a considerable asymmetry between the propagation velocities, whereas lower values of $\mu$ result in a negligible asymmetry. Finally, we note that an energy offset of the Dirac cones with respect to the Fermi level is only present for zigzag ribbons.}
\label{bands}
\end{figure}

Tight-binding numerical simulations of graphene nanoribbons corresponding to the system described by the Hamiltonian \eqref{hardwall} were conducted for both armchair and zigzag prinstine interfaces using the Kwant code.\cite{Groth2014Kwant} We have used a set of unrealistic parameters that emphasize the actual phenomenology and keep a lower computational cost. For the corresponding results obtained from realistic parameters we refer the reader to \cite{Lado2015Majorana, Young2012Spin, Kharitonov2012Edge, Young2014Tunable, Lado2014Noncollinear}. The complete electronic band structure and low-energy spectrum for $\theta=\pi$ are shown in Fig. \ref{bands}. Clearly, different phenomenologies are expected for armchair and zigzag interfaces. For zigzag boundaries, Fig. \ref{bands} (d) and (e), there is an energy offset for the Dirac cones in relation to the Fermi level, corresponding to the different $\mu_1$ and $\mu_2$ in the phenomenological model \eqref{hlado}. On the other hand, for armchair boundaries, Fig. \ref{bands} (a), (b) and (c), there is a gap, captured by the energy $w$ in \eqref{hlado}, due to intervalley scattering. This gap can be softened by considering smooth functions, such as $\tanh (y)$ at the interfaces, instead of Heaviside step functions of \eqref{hardwall}.

Next, we consider the effect of varying the chemical potential at the superconducting regions. For higher values, see Fig. \ref{bands} (b) for armchair interfaces and (d) for zigzag boundaries, there is an obvious difference in the propagation velocities of the chiral modes, captured by the different $v_1$ and $v_2$ in \eqref{hlado}. This is probably because, at the high-doping regime, we are sufficiently far away from the Fermi level. In this case, the linear expansion used to derive the continuum model for graphene, on which the Hamiltonian \eqref{hardwall} is based, does not reproduce all relevant phenomena.\cite{PhysRevB.29.1685} In other words, higher-order terms in momentum must be taken into account. Thus, for simplicity, we restrict our analysis to the commonly used low-doping regime,\cite{Titov2006Josephson,PhysRevB.29.1685} in which $v_1 \approx v_2$, see Fig. \ref{bands} (c) and (e). 

Therefore, we expect that termination physics may account for the remaining terms in the Hamiltonian. That is what we derive next.

\subsection{Analytical treatment}

The effect of termination physics may be addressed by including in the Hamiltonian effective potentials that enforce the desired boundary conditions. Since we have already accounted for Andreev reflections by imposing the continuity of the wave functions at the interface, we need only consider effective potentials describing normal reflections. 

Consider a general energy independent boundary condition for the Dirac equation corresponding to the following linear restriction on the wave function
\begin{align}
	\psi(\vect{r}_B)=M\psi(\vect{r}_B),
\end{align}
where $M$ is an arbitrary Hermitian and unitary matrix. It can be expressed in the form of an additional confinement potential at the boundary $\vect{r}_B$ as:\citep{McCann2004Symmetry}
\begin{align}\label{cpot}
V_{term}(\vect{r}_B) = v_{term} \delta(\vect{r}_B) \tilde{M},
\end{align}
where $v_{term}$ is a constant that represents the strength of the potential. The relation between the matrices $M$ and $\tilde{M}$ can be easily obtained by integrating the Dirac equation including the confinement potential \eqref{cpot} across an infinitesimal width of the boundary, leading to 
\begin{align}
\tilde{M} = -\frac{i}{v} \vect{J} M,
\end{align}
where $\vect{J}$ is the current operator. 

For normal reflections,\cite{Akhmerov2008Boundary} there are three contributions to $\tilde{M}$,
\begin{align}
\tilde{M}_{ac} = \tau_3 \otimes \rho_1 \otimes s_3 \otimes \sigma_0,\\
\tilde{M}_{im} = \tau_3 \otimes \rho_3 \otimes s_3 \otimes \sigma_0, \\
\tilde{M}_{zz} = \tau_3 \otimes \rho_3 \otimes s_1 \otimes \sigma_0,
\end{align}
corresponding to armchair (\textit{ac}), infinite mass (\textit{im}, sublattice imbalance) and zigzag (\textit{zz}) potentials. So that the effective potential enforcing normal reflections has the following form:
\begin{align}\label{term_pot}
V_{term}(y)& = V_{term}^{zz}(y) + V_{term}^{im}(y) + V_{term}^{ac}(y).
\end{align}

We include the effect of the termination potential \eqref{term_pot} in the Hamiltonian as
\begin{align}\label{add_term}
\mathcal{H}_{\text{eff}}^{\alpha \beta} \mapsto \mathfrak{H}_{\text{eff}}^{\alpha \beta} = \mathcal{H}_{\text{eff}}^{\alpha \beta} + \langle \psi_{\alpha} | V_{term}(y) | \psi_{\beta} \rangle.
\end{align}
Noting also that the $y$-dependence of the $V_{term}$ can be neglected, since the form of the wave functions $\psi_{\alpha}(y)$ guarantees that such terms have support only at the interface, the resulting Hamiltonian, in the same representation of Eq. \ref{heff1}, reads
\begin{align}\label{final_eff}
\mathfrak{H}_{\text{eff}}=\left(
\begin{array}{cccc}
\mu_1 + \tilde{v} p & b_{\theta} & w & 0 \\
b_{\theta} & \mu_2 - \tilde{v} p & 0 & w \\
w & 0 & - \mu_2 - \tilde{v} p & b_{\theta}\\
0 & w & b_{\theta} & -\mu_1 + \tilde{v} p
\end{array}\right).
\end{align}
The different chemical potentials $\mu_1$ and $\mu_2$ derive from the infinite mass and zigzag potentials, and the intervalley mixing energy $w$, from the armchair potential. The explicit expressions for the parameters are:
\begin{subequations}
\label{cumbersome}
\begin{align}\label{cumbersome1}
\tilde{v} &= \frac{1}{\mathcal{N}^2} \left(2 \int_{-\infty}^0 dy\  \chi(y)  -\frac{2 \Delta }{\Delta ^2+\mu^2}\right), \displaybreak[3] \\
\label{cumbersome2}
\tilde{\Delta} &= \frac{B}{\mathcal{N}^2} \int_{\infty}^0 dy\ \chi(y), \displaybreak[3] \\
\label{cumbersome3}
\chi(y)&=\frac{4 m e^{y \sqrt{B^2 y^2+4 m^2}}}{\sqrt{B^2 y^2+4m^2}}, \displaybreak[3] \\
\label{cumbersome4}
\tilde{\mu} &= \frac{B v_{im}}{\mathcal{N}^2} \int_{-\infty}^0 dy\ \frac{y \chi(y)}{m}, \displaybreak[3] \\
\label{cumbersome5} 
\delta \tilde{\mu} &= \frac{2}{\mathcal{N}^2}\left( v_{zz} \int_{-\infty}^0 dy\ \chi(y) + \frac{\Delta  v_{zz} +\mu v_{im}}{\Delta ^2+\mu^2}\right), \displaybreak[3] \\
\label{cumbersome6}
w &= \frac{2 v_{ac}}{\mathcal{N}^2}\left(\frac{\Delta }{\Delta ^2+\mu^2} + 2 \int_{-\infty}^0 dy\ e^{y \sqrt{B^2 y^2+4 m^2}} \right),\\
\label{cumbersome7}
\mu_1&=\tilde{\mu}+\delta\tilde{\mu},\\
\label{cumbersome8}
\mu_2&=\tilde{\mu}-\delta\tilde{\mu}.
\end{align}
\end{subequations}

Thus, the effective Hamiltonian \eqref{final_eff} completely describes the expected phenomenology of graphene terminations. The chemical potentials $\mu_1$ and $\mu_2$ shift the Dirac cones for zigzag interfaces. On the other hand, the armchair potential does not shift the cones, but couple different valleys, leading to a gap opening energy $w$. We can finally check the existence of Majorana zero modes for finite systems and explore the topological classification.

\section{Majorana modes and the topological classification}
\label{majorana_modes}

\subsection{Majorana modes}

For the sake of clarity, we will use in this section the following spinor representation $\psi = (\psi_{++}, \psi_{+-}, \psi_{-+}, \psi_{--})^T$. In this representation, the Hamiltonian \eqref{heff1} takes the simpler form
\begin{align}\label{majorana_ham}
\tilde{\mathcal{H}}_{\text{eff}} = v p \kappa_3 \otimes \eta_3 + b_{\theta} \kappa_0 \otimes \eta_1,
\end{align}
where $\{\kappa_{\alpha}\}_{\alpha=0}^3$ and $\{\eta_{\alpha}\}_{\alpha=0}^3$  are sets with the identity and Pauli matrices in the usual representation. The indices of the spinor components, $\psi_{\kappa \eta}$, represent the eigenvalues of $\kappa_3$ and $\eta_3$, respectively.

We can now show that gap closings in \eqref{majorana_ham} result in zero energy states that are, indeed, Majorana modes. Making the $x$-dependence of the intravalley coupling explicit, \emph{i.e.}, $b_{\theta}=b_{\theta}(x)$, and assuming that it changes sign at $x=0$, we can expand \eqref{majorana_ham} around $x=0$ to obtain:
\begin{align}
\tilde{\mathcal{H}}_{\text{eff}}^2 = \tilde{v}^2 p^2 \kappa_0 \otimes \eta_0 + x^2 \mathcal{B}_{\theta}^2 \kappa_0 \otimes \eta_0 - \mathcal{B}_{\theta}\tilde{v}\kappa_3 \otimes \eta_2,
\end{align}
with $\mathcal{B}_{\theta} = \partial_x b_{\theta}|_{x=0}$. The energy spectrum is then easily obtained:
\begin{align}
E_n^{\kappa, \eta} = \pm \sqrt{ 2\mathcal{B}_{\theta}\tilde{v}\left(n+\frac{1}{2}\right) - \kappa \eta \mathcal{B}_{\theta}\tilde{v}}
\end{align}
with $\kappa,\eta=\pm$. Also, note that the ground state is doubly degenerate:
\begin{align}
\gamma_{\pm}(x) = \langle x | n=0, \kappa = \pm, \eta = \pm \rangle.
\end{align}

The appearance of these bound states should not be surprising, since it corresponds to a change of the topological invariant on the phenomenological model \eqref{hlado},\cite{Lado2015Majorana} for $\mu_1=\mu_2=w=0$ and $v_1=v_2=\tilde{v}$ as we see below.

\subsection{Topological classification}

Finally, we consider how the extra terms of the effective Hamiltonian \eqref{final_eff} affect the emergence of Majorana modes. Using the same representation of \eqref{majorana_ham}, the effective Hamiltonian \eqref{final_eff} reads:
\begin{align}\label{eff-equal-ferr}
\mathfrak{H}_{\text{eff}} &= \tilde{\mu}\ \kappa_3 \otimes \eta_0 + \delta \tilde{\mu}\ \kappa_0 \otimes \eta_1 + w\ \tau_1 \otimes \eta_0 \nonumber \\
&+ \tilde{v}p_x\ \tau_3 \otimes \eta_1 - b_{\theta} \kappa_0 \otimes \eta_3.
\end{align}
Interestingly, if we compare \eqref{eff-equal-ferr} to the Hamiltonian of a topological superconducting ferromagnetic nanowire:\cite{Dumitrescu2015Majorana}
\begin{align}\label{ferro}
H_{\text{ferr}} &= [t p_x^2 - (\mu+2t)] \sigma_0 \otimes \tau_3 + [\Delta_s\ \sigma_0 + p_x \vect{d} \cdot \vect{\sigma}] \otimes \tau_1 \nonumber \\ &+ \vect{V}\cdot \vect{\sigma} \otimes \tau_0,
\end{align}
where $t$ is the hopping constant, $\mu$ is the chemical potential, $\Delta_s$($\vect{d}$) is an $s$($p$)-wave superconducting order parameter, $\vect{V}$ is the Zeeman term and the matrices $\tau_{\alpha}$ and $\sigma_{\alpha}$ designate particle-hole and spin spaces, respectively, we find the following correspondence:
\begin{gather}
t=d_1=d_2=V_2=0,\\
\mu \leftrightarrow b_{\theta},\\
\Delta_s \leftrightarrow \delta \tilde{\mu},\\
d_3 \leftrightarrow \tilde{v},\\
V_1 \leftrightarrow w,\\
V_3 \leftrightarrow \tilde{\mu}.
\end{gather}

Thus, we can conclude that the system described by the effective Hamiltonian \eqref{eff-equal-ferr} has a charge-conjugation-like symmetry, described by some anti-unitary operator that anticommutes with the Hamiltonian.\cite{Lado2015Majorana} On the other hand, $w$ and $\tilde{\mu}$ explicitly break any time-reversal-like symmetry, described by anti-unitary operators that commute with the Hamiltonian and square to $-\mathbb{1}$.\cite{Dumitrescu2015Hidden, Dumitrescu2015Majorana} Finally, there is a pseudo-time-reversal-like symmetry, described by some anti-unitary operator that commutes with the Hamiltonian and squares to $+\mathbb{1}$, that is explicitly broken for non-zero $\tilde{\mu}$.\cite{Dumitrescu2015Hidden, Dumitrescu2015Majorana, Manesco2018Hidden} Therefore, for disordered interfaces, corresponding to non-zero $w$ and $\tilde{\mu}$, only the charge-conjugation-like symmetry is preserved and the system is in the $D$ class. \cite{Lado2015Majorana}

\section{Conclusion}
\label{conclusions}

In this paper, we presented a formal derivation of the phenomenological Hamiltonian proposed to describe graphene-superconductor junctions at low-doping. Our approach allows the understanding of such systems in terms of experimentally controllable parameters, although the correspondence between the phenomenological and experimental parameters is highly nontrivial. It was found that, in order to completely describe the low-energy spectrum of such junctions, effects related to normal reflection must be taken into account. Generalizing our results to describe graphene-superconductor junctions also at high doping requires considering higher order corrections in the continuum model for graphene.

Finally, we demonstrated the emergence of Majorana zero modes in the system and provided its topological classification by mapping the effective Hamiltonian \eqref{final_eff} to the Hamiltonian describing a superconducting ferromagnetic nanowire. Unfortunately, the presence of interface potentials breaks all discrete symmetries but charge conjugation. Hence, there is only one non-trivial topological phase possible corresponding to the class D.

\section*{Acknowledgments}
The work of ALRM was supported by São Paulo Research Foundation Grant No. 2016/10167-8. DRJ is a CNPq researcher. The authors thank A. Akhmerov and P. San-Jose for helping us with the numerical implementation. We also thank J. Lado for very useful discussions.

\appendix

\section{Gamma matrices}
\label{gamma}

The Hamiltonian \eqref{hardwall} is written in the valley-symmetric representation $\psi = (c, \mathcal{T}c)^T$, where:
\begin{align}
c&=
\left(
\begin{array}{c}
 c_{KA\uparrow} \\
 c_{KA\downarrow} \\
 c_{KB\uparrow} \\
 c_{KB\downarrow} \\
 c_{K^*B\uparrow} \\
 c_{K^*B\downarrow} \\
 c_{K^*A\uparrow} \\
 c_{K^*A\downarrow} \\
\end{array}
\right),\quad \mathcal{T} = -i \rho_2 \otimes s_2 \otimes \sigma_2 \mathcal{K}.
\end{align}
The $\Gamma$-matrices are defined as:
\begin{align}
\Gamma_0 &:= \tau_3 \otimes \rho_0 \otimes s_0 \otimes \sigma_0, \\
\Gamma_1 &:= \tau_3 \otimes \rho_0 \otimes s_1 \otimes \sigma_0, \\
\Gamma_2 &:= \tau_3 \otimes \rho_0 \otimes s_2 \otimes \sigma_0, \\
\Gamma_3 &:= \tau_0 \otimes \rho_3 \otimes s_3 \otimes \sigma_1, \\
\Gamma_4 &:= \tau_0 \otimes \rho_0 \otimes s_0 \otimes \sigma_3, \\
\Gamma_5 &:= \tau_1 \otimes \rho_0 \otimes s_0 \otimes \sigma_0,
\end{align}
where $\{\tau_{\nu}\}$ corresponds to electron-hole ($c$ and $\mathcal{T}c$), $\{\rho_{\nu}\}$ to valley, $\{s_{\nu}\}$ to sublattice and $\{\sigma_{\nu}\}$ to spin degrees of freedom, respectively. The index $\nu=0$ corresponds to the identity and $\nu=1,2,3$, to the three Pauli matrices in the usual representation.

\section{Interface modes}
\label{eigenfunctions}

The construction of a basis for the solutions of quantum Hall antiferromagnetic graphene-superconductor junctions outlined in Sec. \ref{model} leads to the following set of eigenspinors $\{\psi_{\kappa \eta}\}$, $\kappa,\eta=\pm$, in which the upper indices $<$ and $>$ indicate the regions $y<0$ and $y>0$, respectively. Here, to avoid cluttering, we introduce the functions:
\begin{align}
\phi_1(B,m,y)&=\frac{2 m e^{\frac{1}{2} y \sqrt{B^2 y^2+4 m^2}}}{\sqrt{B y \left(B y-\sqrt{B^2 y^2+4 m^2}\right)+4 m^2}},\\
\phi_2(\Delta, \mu, y) &= e^{-\Delta y} \sin (\mu  y),\\
\phi_3(\Delta, \mu, y) &= e^{-\Delta y} \cos (\mu  y),
\end{align}
so that the eigenspinor basis can be written as
\begin{align}
\psi_{++}^<(y)&=\frac{1}{\mathcal{N}}
\left(
\begin{array}{c}
 0 \\
 0 \\
 0 \\
 0 \\
 0 \\
 \phi_1(B,m,y) \\
 -\phi_1(B,m,y) \\
 0 \\
 0 \\
 0 \\
 0 \\
 0 \\
 \phi_1(B,m,y) \\
 0 \\
 0 \\
 \phi_1(B,m,y) \\
\end{array}
\right),
\end{align}

\begin{align}
\psi_{++}^>(y)&=\frac{1}{\mathcal{N}}
\left(
\begin{array}{c}
 0 \\
 0 \\
 0 \\
 0 \\
 -\phi_2(\Delta, \mu, y)  \\
 \phi_3(\Delta, \mu, y) \\
 -\phi_3(\Delta, \mu, y) \\
 -\phi_2(\Delta, \mu, y)  \\
 0 \\
 0 \\
 0 \\
 0 \\
 \phi_3(\Delta, \mu, y) \\
 \phi_2(\Delta, \mu, y)  \\
 -\phi_2(\Delta, \mu, y)  \\
 \phi_3(\Delta, \mu, y) \\
\end{array}
\right),
\end{align}

\begin{align}
\psi_{+-}^<(y)&=\frac{1}{\mathcal{N}}
\left(
\begin{array}{c}
 0 \\
 0 \\
 0 \\
 0 \\
 \phi_1(B,m,y) \\
 0 \\
 0 \\
 -\phi_1(B,m,y) \\
 0 \\
 0 \\
 0 \\
 0 \\
 0 \\
 \phi_1(B,m,y) \\
 \phi_1(B,m,y) \\
 0 \\
\end{array}
\right),
\end{align}

\begin{align}
\psi_{+-}^>(y)&=\frac{1}{\mathcal{N}}
\left(
\begin{array}{c}
 0 \\
 0 \\
 0 \\
 0 \\
 \phi_3(\Delta, \mu, y) \\
 -\phi_2(\Delta, \mu, y) \\
 -\phi_2(\Delta, \mu, y) \\
 -\phi_3(\Delta, \mu, y) \\
 0 \\
 0 \\
 0 \\
 0 \\
 \phi_2(\Delta, \mu, y) \\
 \phi_3(\Delta, \mu, y) \\
 \phi_3(\Delta, \mu, y) \\
 -\phi_2(\Delta, \mu, y) \\
\end{array}
\right),
\end{align}

\begin{align}
\psi_{-+}^<(y)&=\frac{1}{\mathcal{N}}
\left(
\begin{array}{c}
 0 \\
 \phi_1(B,m,y) \\
 \phi_1(B,m,y) \\
 0 \\
 0 \\
 0 \\
 0 \\
 0 \\
 -\phi_1(B,m,y) \\
 0 \\
 0 \\
 \phi_1(B,m,y) \\
 0 \\
 0 \\
 0 \\
 0 \\
\end{array}
\right),
\end{align}

\begin{align}
\psi_{-+}^>(y)&=\frac{1}{\mathcal{N}}
\left(
\begin{array}{c}
 \phi_2(\Delta, \mu, y) \\
 \phi_3(\Delta, \mu, y) \\
 \phi_3(\Delta, \mu, y) \\
 -\phi_2(\Delta, \mu, y) \\
 0 \\
 0 \\
 0 \\
 0 \\
 -\phi_3(\Delta, \mu, y) \\
 \phi_2(\Delta, \mu, y) \\
 \phi_2(\Delta, \mu, y) \\
 \phi_3(\Delta, \mu, y) \\
 0 \\
 0 \\
 0 \\
 0 \\
\end{array}
\right),
\end{align}

\begin{align}
\psi_{--}^<(y)&=\frac{1}{\mathcal{N}}
\left(
\begin{array}{c}
 \phi_1(B,m,y) \\
 0 \\
 0 \\
 \phi_1(B,m,y) \\
 0 \\
 0 \\
 0 \\
 0 \\
 0 \\
 -\phi_1(B,m,y) \\
 \phi_1(B,m,y) \\
 0 \\
 0 \\
 0 \\
 0 \\
 0 \\
\end{array}
\right),
\end{align}

\begin{align}
\psi_{--}^>(y)&=\frac{1}{\mathcal{N}}
\left(
\begin{array}{c}
 \phi_3(\Delta, \mu, y) \\
 \phi_2(\Delta, \mu, y) \\
 -\phi_2(\Delta, \mu, y) \\
 \phi_3(\Delta, \mu, y) \\
 0 \\
 0 \\
 0 \\
 0 \\
 \phi_2(\Delta, \mu, y) \\
 -\phi_3(\Delta, \mu, y) \\
 \phi_3(\Delta, \mu, y) \\
 \phi_2(\Delta, \mu, y) \\
 0 \\
 0 \\
 0 \\
 0 \\
\end{array}
\right).
\end{align}

We end this appendix with some important observations regarding the spinor structure of the above basis. First, we note that in the QHAF region the well known identification of valley and sublattice degrees of freedom at the zeroth Landau level still holds, as well as the well defined spin polarization of these degrees of freedom. Indeed, this suggests that $\kappa$ is related to valley polarization, whereas $\eta$ is related to the helicity of the modes. On the superconducting side, although both spin polarizations are present, they are described by orthogonal functions. Thus, exactly at the interface, corresponding to $\theta=\pi$, only the chiral propagating modes remain. Finally, the presence of superconductivity adds the charge-conjugation symmetry. So that by changing the canting angle, the chiral superconducting modes become gapped and the system can be regarded as an one-dimensional topological superconductor, as the effective Hamiltonian \eqref{final_eff} suggests.

\bibliography{graphene-SC}

\end{document}